\documentclass{egpubl}
\usepackage{envirvis2023}

 \WsSubmission    % uncomment for submission to EuroVis Workshop
\electronicVersion % can be used both for the printed and electronic version

\ifpdf \usepackage[pdftex]{graphicx} \pdfcompresslevel=9
\else \usepackage[dvips]{graphicx} \fi
\usepackage{lineno}

\PrintedOrElectronic

\usepackage{t1enc,dfadobe}

\usepackage{egweblnk}
\usepackage{cite}

\usepackage{xcolor}

\title[An IDSS for Analyzing Time Related Restrictions in Renaturation and Redevelopment Planning Projects]%
      {An Interactive Decision Support System for Analyzing Time Related Restrictions in Renaturation and Redevelopment Planning Projects}

\author[Y. Annanias, C. Meinecke \& D. Wiegreffe]{Y. Annanias$^{1}$\orcid{0000-0003-2259-1254}, C. Meinecke$^{1}$\orcid{0000-0002-5637-9975} and D. Wiegreffe$^{1}$\orcid{0000-0002-5356-6732} \\
$^1$Leipzig University, Image and Signal Processing Group, Germany
}

\usepackage{enumitem}
\usepackage{overpic}
\begin{document}

%\linenumbers

% \teaser{
%  \includegraphics[width=\linewidth]{eg_new}
%  \centering
%   \caption{New EG Logo}
% \label{fig:teaser}
% }

\maketitle
%memory - knowledge
\begin{abstract}
% Motivation
The operation of open-cast lignite mines is a large intervention in nature, making the areas uninhabitable even after closing the mines without renaturation processes.
Renaturation of these large areas requires a regional planning process which is tied to many conditions and restrictions, such as environmental protection laws.
% Problem
The related information is available only as unstructured text in a variety of documents.
Associated temporal aspects and the geographical borders to these textual information have to be linked manually so far.
This process is highly time-consuming, error-prone, and tedious. 
Therefore, the knowledge of experts is often used, but this does not necessarily include all the relevant information. 
% Lösung
In this paper, we present a system to support the experts in decision-making of urban planning, renaturation, and redevelopment projects. 
The system allows to plan new projects, while considering spatial and temporal restrictions extracted from text documents.
With this, our presented system can also be used to verify compliance with certain legal regulations, such as nature conservation laws.

%to create new polygons , such as bike trails, , like bird breeding or tree pruning periods. 
%With this, our presented system can also be used to verify compliance with certain legal and nature conservation regulations.

%\begin{classification} % according to https://dl.acm.org/ccs

%\end{classification}

\end{abstract}

%-------------------------------------------------------------------------
%4 to 8 pages including references

\section{Introduction}
%Problem /Motivation
Regional planning processes are complex and often require a large amount of data of different types. 
%This is especially true for the renaturation of larger areas, but also for urban planning and redevelopment. 
This is particularly applicable to the renaturation of larger areas and also holds true for urban planning and redevelopment.
The data generated by the expert reports must then be linked to additional data such as sensor data or time points in order to be able to draw conclusions. %sensor konkretter z.b. wetter?
For example, there are restrictions that can lead to delays because certain activities are not permitted by law at certain times.
These include, for example, periods where tree pruning is not allowed, or various bird nesting periods. 
During bird nesting periods, any activity that would disturb the animals is prohibited. % during these times. 
Therefore, knowing the specific restrictions, checking them, and being able to communicate them to others, such as project managers and the public, is essential for project planning.

% baumschnitt, allergie-pollen?, mehr auf bauaspekt eingehen
% environmental research, decision-making and public outreach.
% Land use research

% Project + erweiterung
In a previous work in the context of regional planning processes, we developed an interactive decision support system that works with large amounts of geographical data and takes care of the representation and analysis of overlapping areas \cite{YA_IDS}.
The system also supports the analysis and linkage of weather-related restrictions, such as the amount of precipitation per area and weather warnings~\cite{annanias2022interactive}.

\begin{figure*}[tb]
	\centering
	 \begin{overpic}[width=1\linewidth]{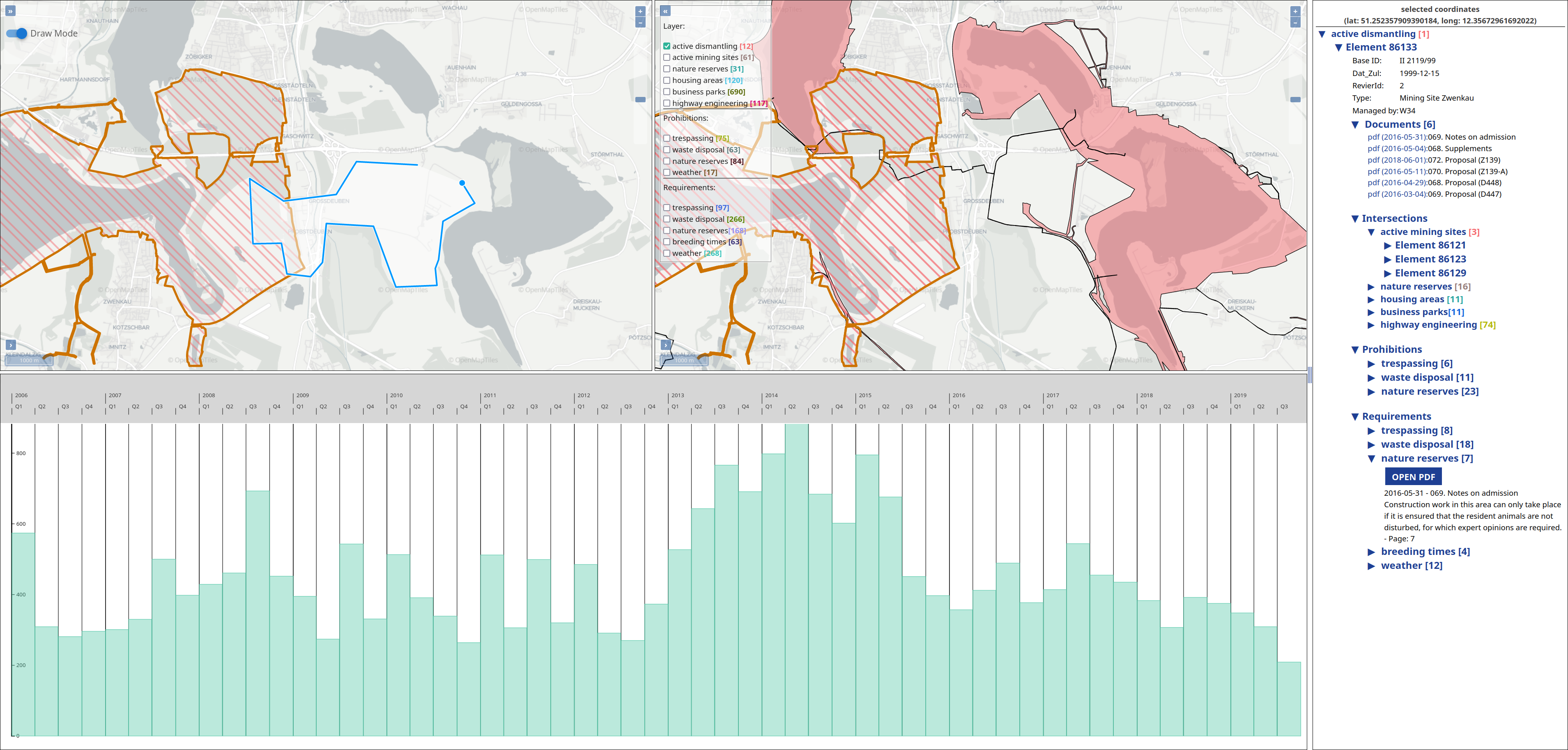} 
     	 \put(2,46.4){\textbf{\textit{a)}}}
     	 \put(44,46.4){\textbf{\textit{b)}}}
     	 \put(84,46.4){\textbf{\textit{c)}}}
     	 \put(2,22.4){\textbf{\textit{d)}}}
     \end{overpic}
	\caption[Component]{\label{fig:mainTool}
		 The main decision support system with multiple views.
		 (a) and (b) are two linked maps showing the same section.
		 In (b), polygons of the category ``active dismantling'' are displayed, whereby one was selected (orange border).
		 By brushing, this polygon is also displayed in (a). 
		 In (a), a new project is planned by drawing a new polygon. 
		 (c) is the information panel, which lists important documents and their contents, as well as a list of all polygons that overlap with the selected one. 
		 (d) is a timeline visualization that shows how many documents are available in the displayed area.
	}
\end{figure*}
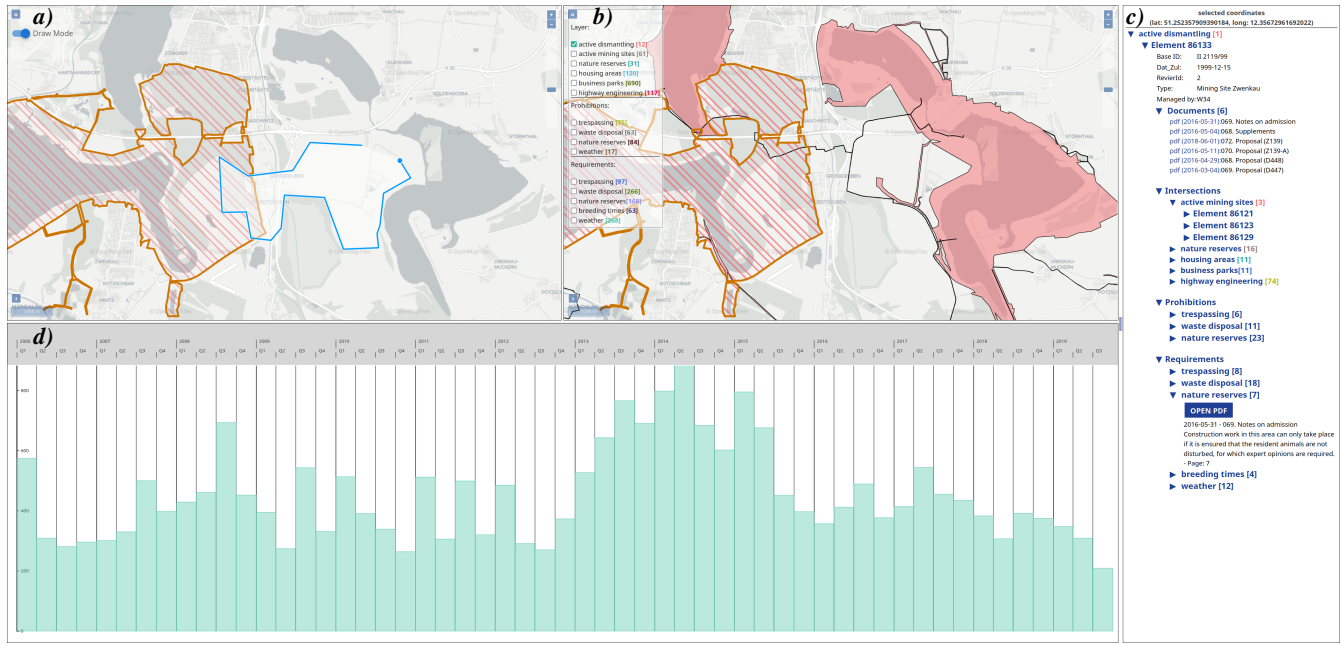

%Was machen wir/visualiseren wir - Lösung
Here, we extend this work to the aspect of text documents that have temporal restrictions, such as breeding times, and show their influence on the planning of the execution of construction processes. 
For this purpose, we digitally process the text documents and list the corresponding restrictions.
Then, we support the search for corresponding documents with such restrictions by a timeline visualization. 
In addition, new projects can be planned that keep track of and display existing restrictions.
For example, a new bicycle path can be planned by drawing a path or a new area on the map. 
Figure \ref{fig:mainTool} provides an overview of the system and the extensions.
%polygon oder lieber path -> punkte mit radius setzen wäre einfacher als polygon zeichnen?

% Project Partner
Our project partner is the Lausitzer und Mitteldeutsche Bergbauverwaltungsgesellschaft mbh (LMBV), which is mainly concerned with the renaturation of open-cast mining areas in the eastern part of Germany. 
Landscape redesign planning and implementation plays a major role in this process, which involves the management of large amounts of georeferenced text data. 
Part of the renaturation is the recultivation of the flora and fauna in the affected areas, but also the settlement of residential areas and the provision of industrial areas.

Our contributions are summarized as follows:
\begin{itemize}
    \item Support in searching for temporal restrictions in text documents (timeline view)
    \item Support in the planning of new projects while taking into account such restrictions (polygon drawing)
    \item Support in monitoring existing projects and checking for compliance with the restrictions (brushing and linkage of views)
\end{itemize}

This introduction is followed by a description of the data and the necessary preprocessing steps that will be used to fulfill the tasks and the resulting requirements for the system described afterwards.
This overview is followed by related works with a focus on geospatial and temporal visualization systems.
Next, we describe our implemented solution and give an example use case. 
Finally, limitations are considered, and a conclusion is drawn.
%\section{Background and Motivation}\label{sec:currentStateReq}
\section{Data \& Data Preprocessing}
As part of our collaboration, the LMBV provided us with a test data set.
This data set consists mainly of more than 60 gigabytes of text documents with georeferences, including reports on the general condition of the areas and the current forest stand, legal opinions and planning requirements, soil surveys, and listings of animal and plant species typically found in the region.
The georeferences are provided as point and area descriptions through polygon data.

\subsection{Information Extraction}
These documents are prepared by the experts of the LMBV or were created by external experts on behalf of the LMBV in advance of planning processes and contain, among other things, legal restrictions and other limitations that must be adhered to. 
In order to be able to find restrictions regarding bird breeding times without having to read all the documents completely, digital processing of the documents is necessary. 
In a previous work, Schröder et al. \cite{schroder2021mining} processed these data using optical character recognition (OCR) and a subsequent classification of the extracted sentences using an active learning approach.
The classification divides the extracted restrictions into ``Prohibitions'' (to refrain from an action in general) and ``Requirements'' (to refrain from an action under certain conditions). 

This approach can be easily adapted by means of suitable keywords and a new training, to search for restrictions concerning the topic of breeding times.
Accordingly, this type of restriction falls into the category of ``Requirements'' because only certain actions are prohibited in a certain period of time.
In this way, we obtain all documents and, respectively, all sentences from the documents that match the topic.

However, some of these restrictions are valid only within a specified time frame.
This information is most often contained directly in the extracted sentences. 
Using a rule-based named entity recognition (NER), the time data was also extracted and stored with reference to the sentence. 
The rules were based on patterns such as date specifications (\textit{DD.MM.YYYY}), mention of a month surrounded by numbers (\textit{February 2023}), or other typical terms for dates (\textit{year 2021}), as well as a variety of combinations thereof.
Since not all sentences contain a direct reference to the appropriate time period, the sentences had to be manually annotated with a time reference.

All documents are linked to the areas where these restrictions apply, which means that all documents are geographically referenced.
Therefore, all areas are available as additional polygons, which can be displayed on a map using a geographic information system (GIS).

\subsection{Data Linkage}
In order to quickly find the data and keep the response times of the database short, the data must be efficiently linked and stored. 
In our previous work, we used a Neo4j graph database for this, where polygons and documents are each stored as nodes and linked to each other via an edge.
Additionally, the polygons are intersected with each other, and if an overlap exists, both polygon nodes are connected via an edge. 
This is useful because the restrictions of the documents of both polygons apply equally in the overlapping area. 
In addition, in the case of a new area, the overlaps can be used to determine what kind of restrictions are expected.

\begin{figure}[tb]
	\centering
	\includegraphics[width=1\linewidth]{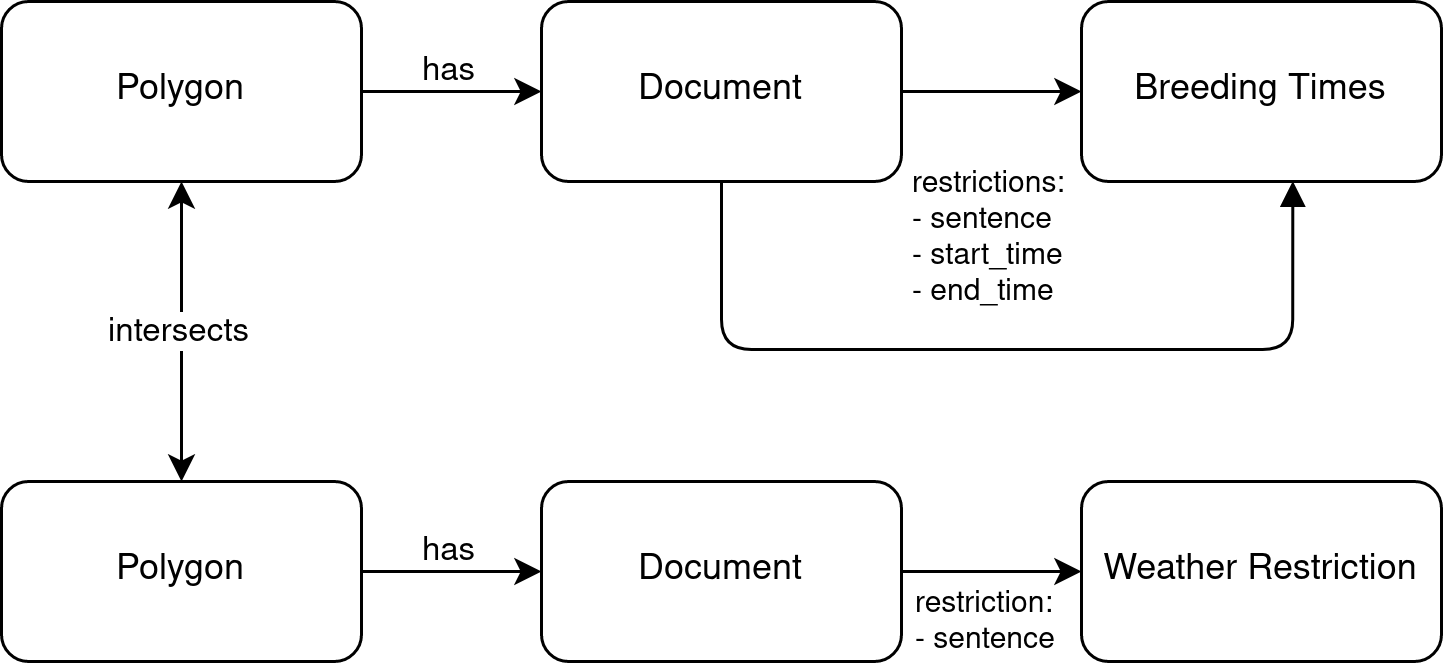}
	\caption[Component]{\label{fig:graph}
		 A section of the graph data model showing two polygon nodes, each with a document node, and two restriction type nodes. 
		 The upper document node contains two sentences regarding breeding times, the attributes such as the corresponding sentence as well as the time points of this restriction are stored at the corresponding edge.
	}
\end{figure}

%TODO den Satz umschreiben?
Similarly, each restriction class is represented by a node, so for our current problem, the node ``Breeding Times'' was added (see Figure \ref{fig:graph}). 
Each extracted and classified sentence that refers to a breeding time is created as an edge between the document in which it is located and the breeding time node. 
Such an edge contains the associated sentence, as well as the start and end time for the time constraint, as attributes.
Thus, a query with the node \textit{Breeding Times} immediately returns all areas for which such a restriction exists; at the same time, the result set can also be filtered by geographical location.
Then, the result can also be sorted and interpreted accordingly via the time attributes at the edges of the sentences.

%TODO figure 2
%    - runde ecken
%    - pfeile mit rundung
%    - pfeile mit fläche
%    - eine farbe?
%    - mehr abstand zwischen den letzen boxen, dafür labels zweizeilig
%    - box shadow?
%Begründen über Tasks -> haben wir folgende Requirements
\section{Tasks \& Requirements}
\subsection{Task}
A typical domain-specific task is to determine what temporal constraints exist in a specific area under study and to check whether they are valid for a given point in time.
Currently, restrictions have to be manually searched in documents, which is tedious and time consuming.
In some cases, the experts' experiences and knowledge are also used, which can mean that some restrictions are not taken into account. 
In particular, experts who are new to the company cannot draw on this knowledge.

When planning new projects, it is also interesting to include existing projects within the same area. 
Because the restrictions that apply to one area automatically apply in the overlapping section of the other area.
In this way, problems that can occur or that are to be expected due to legal restrictions can be examined in advance.

\subsection{Requirements}
In consultation with our experts, it is necessary that the resulting system satisfies the following requirements to support domain-specific tasks and facilitate planning processes:
\begin{itemize}[itemindent=0.25cm]
    \item [\textbf{R1}]\label{req:R1} Time restrictions in documents must be found efficiently and related documents should be listed. 
    \item [\textbf{R2}]\label{req:R2} The creation of new projects, by drawing polygon areas or paths, should be supported.
    \item [\textbf{R3}]\label{req:R3} For new projects, it must be possible to check whether existing documents contain restrictions that are also valid for the new project.
\end{itemize}

\section{Related Work}
Our system is related to geographic information systems (GIS) used by experts to analyze map-based data and support decisions regarding environmental problems.
For this, we combine geospatial and temporal visualization methods to support decision-making in renaturation processes and urban planning. 
Another related area is the visual analysis of legal documents~\cite{resck2022legalvis,lettieri2017legal}.
%TODO there is little discussion about other research that tackles the specific challenges of extracting and visualizing legal and environmental restrictions in planning processes

\subsection{Geographic Information System}
Geographic information systems (GIS) can provide many ways to measure, interpret, visualize, generalize, and interpolate collected samples \cite{Goodchild:2007}.
Specifically, spatial interpolation techniques are used to complete those spatial locations where measurements are not available or missing. 
GIS have penetrated a variety of fields supporting experts in decision-making and are successfully used to create continuous surfaces in, e.g., meteorology, climatology, criminology, and water management. 
They can combine textual information from the database with spatial data to overcome the difficulties of many spatial decision-making processes. 
The main challenge of GIS has been building a database from unstructured data sets, e.g., reports or media in a variety of data formats, which are now vastly available. 

A recent toolbox of commercial \textit{ARCGIS} software called \textit{LocateXT}~\cite{LocateXT} has enabled connecting a large number of unstructured data sets in different formats (e.g., pdf, ppt, doc) with GIS.
The specific properties of this toolbox include extracting latitude and longitude data from the document, creating a layer based on those, tagging specific properties (i.e., mosque, parks, etc.), data scraping, queries, and more. 
Howell et al.~\cite{Howell:2019} used the toolbox for the parsing, processing, and visualization of scientific articles. 
The tool is not applicable to our project, as it provides point coordinates, and we need area information.
Furthermore, we already have location information from our collaboration partner.
However, the toolbox does not include advanced interaction mechanisms, such as brushing or highlighting. 
Finally, we provide visualization and interaction facilities that are missing from the toolbox.
%We found very limited examples of scientific articles that use the toolbox, as it is only commercially available. 

\subsection{Geovisualization}
Next to GIS, other Geovisualization tools were used for different aspects of environmental data to support decision-making.
An example is visualization in remedial programs, to document, represent, and communicate three-dimensional environmental data to both professionals and the general public~\cite{ling2014environmental}. 
Li et al.~\cite{li2017visual} proposed a visualization tool for the analysis and decision-making of flood risk in cultural heritage using a GIS to show flood peak flow. 
Lei et al. present a geovisualization tool to explore the results of climate simulation in combination with uncertain population growth and water availability based on historical trends in the Niger River basin~\cite{lei2015interactive}.
Lukasczyk et al.~\cite{kolditz2014webgl} use three-dimensional geovisualization to show energy consumption to support decision-making.
Eighausen et al.~\cite{eligehausen2013blendgis} visualize spatial ecology-related data created with \textit{QuantumGIS} in a tool to increase social acceptance in public participation processes for public-funded ecosystem restoration measures and research programs. 
In contrast to these works, we focus on communicating legal restrictions regarding the environment in renaturation and urban planning projects.

\subsection{Visualization of Spatial-Temporal Data}
For renaturation and redevelopment planning projects, the visualization of geospatial and temporal data is important.
For the exploratory analysis of spatio-temporal data, several visualization techniques were proposed, an overview is given by Andrienko et al.~\cite{Andrienko2003}. %provide a list of visualization-based techniques that allow the exploratory analysis of spatio-temporal data. 
\textit{GeoTemCo}~\cite{janicke2013geotemco} allows geospatial and temporal comparison of multiple data sets using a dynamic delaunay triangulation to merge glyphs on the map to avoid visual clutter and was previously applied to data from biodiversity, climatology, and geophysics~\cite{janicke2014utilizing}.
Schlegel et al.~\cite{schlegel2013determining} analyze spatio-temporal patterns of hydrological parameters relevant for flooding.

Outside of environmental science, systems have been proposed for the analysis of spatial-temporal data for crime analysis~\cite{roth2010user} and the exploration and discovery of web information~\cite{dork2008visgets}.
Several works propose spatial-temporal visualizations for social media, news and topic analysis~\cite{li2018visual,sheidin2017time,he2016spatiotemporal,zhang2016visual,kraft2013less}.
Hao et al. visualize customer feedback using sentiment analysis and tag maps~\cite{hao2013visual}, while Reckziegel and Jänicke propose a visualization method for time-varying predominance tag maps~\cite{reckziegel2019time}.
In contrast to these works, our scope lies on legal restrictions in the text documents and not on other aspects such as topics or frequent words.
\section{Visualization and Interaction}
For the design of the visualization system, we follow the Visual Information Seeking Mantra~\cite{shneiderman2003eyes} and the Guidelines for Using Multiple
Views in Information Visualization~\cite{wang2000guidelines}, as this was also applied in the underlying main system. 
Additionally, we refer to the taxonomy or mechanisms for concurrent tool linking (multiple displays showing data) of Andrienko and Andrienko~\cite{Andrienko2006}.
According to their taxonomy, several displays (maps) should show the information for a selected part of the data with important interaction methods for coordination such as highlighting, focusing, zooming, and filtering. 
Brushing can also play an important role in this context. 
They further specify that a change to one display must result in a change to all other linked displays. 
% Baldonado et al. [14] state, that multiple views can reduce cognitive overhead compared to a single more complex view.  In contrast, they indicate that multiple views can have a significant impact on the time and effort required to learn the system. We aim to reduce this by using only uniform views (i.e., same geographical maps). 

\subsection{Tool Overview}
Considering the aforementioned design guidelines, the main system we presented in~\cite{YA_IDS} uses several views (multiple maps and an information panel) arranged in a grid, where all views are linked to each other.
At first, only one map is displayed, which shows the relevant section of a certain region, adjustable by zooming and panning with a panel on the right that lists the associated text data.
If needed, other maps can be added showing the same or a different section, and, if necessary, other data sets, allowing comparison of different data sets with as little disturbing overlap as possible.
To maintain focus, all maps are linked together. 
In this way, the displayed sections of a map can be linked so that zooming and panning have the same effect on all maps.
In addition, the area displayed on one map can be used as a geographic location filter for data selection in other maps and views.
Furthermore, brushing is used to display selected elements in the other views accordingly.

In this work, we added the temporal aspect through a timeline visualization showing temporal restrictions in documents.
Further, we include the possibility to create new projects by drawing polygon areas, while keeping track of existing legal restrictions.
In the same way, additional maps can be attached, and a timeline view can be added, which is responsible for filtering documents according to certain restrictions, categories, and points in time (as shown in Figure~\ref{fig:mainTool}).
All views are arranged according to a grid and resized.
However, the timeline occupies the entire bottom row to have as much space as possible for the visualization.

\begin{figure}[tb]
	\centering
	\begin{overpic}[trim={0 0 0 0},clip, width=\linewidth]{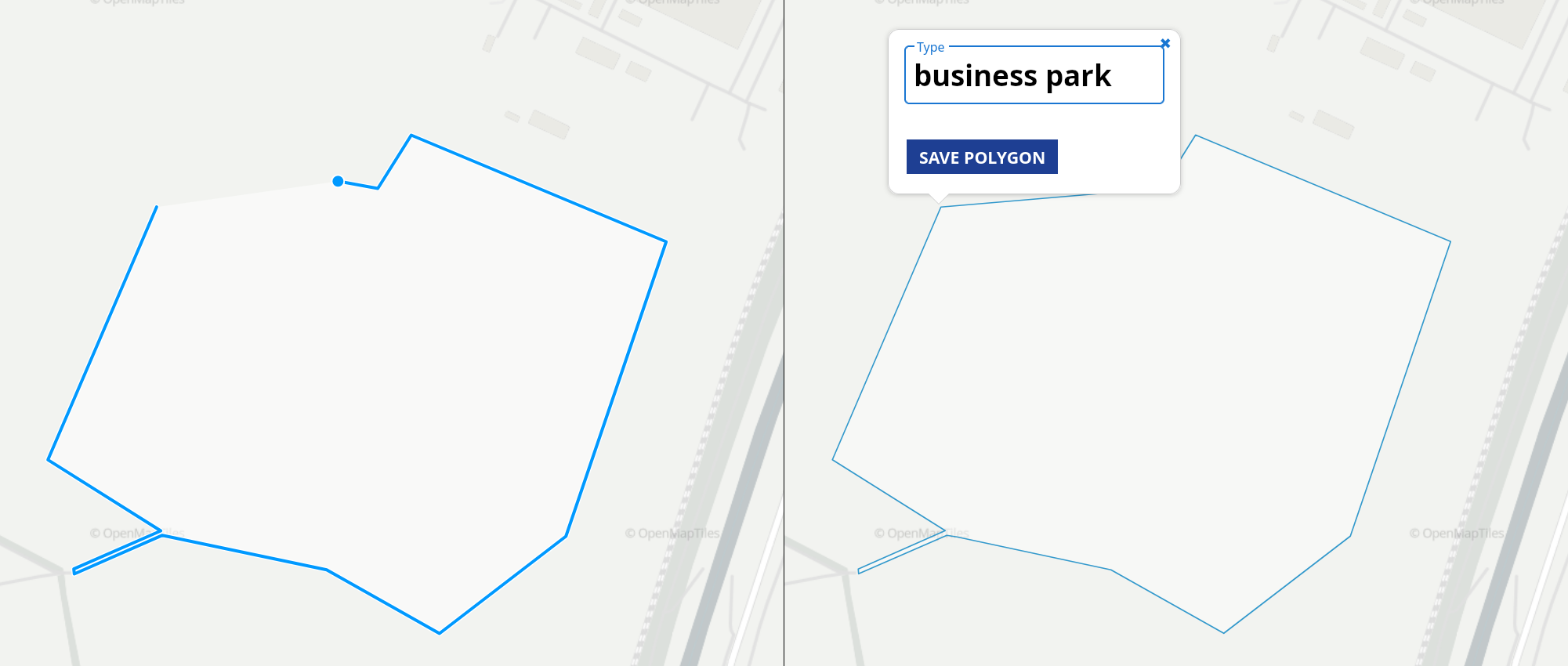}
     	 \put(1,39){\textbf{\textit{a)}}}
     	 \put(51,39){\textbf{\textit{b)}}}
    \end{overpic}
	\caption[Component]{\label{fig:draw}
		 (a) A polygon is drawn directly onto the map, constructing it from individual points. 
		 (b) When the path is closed, the polygon can be saved in a suitable category.
	}
\end{figure}

\subsection{Map}
The polygons are divided into different categories depending on the associated documents.  
Polygons of a category can be displayed independently of each other on the map.
Each category is assigned a color, and the polygons are displayed slightly transparent so that the underlying section of the map remains visible. 
To prevent too much color distortion, the map is only displayed in white and gray tones.
The restriction classes were also used as classes, so that all polygons can be directly displayed that are linked to a document that has restrictions regarding breeding times, for example.

For planning new projects, our extension offers the possibility to draw new polygons directly on the map (Figure~\ref{fig:draw}), which supports \textbf{R2}. 
To do this, the planning expert places one point after the other on the map, which are automatically connected with an edge. 
In this way, the entire polygon can be defined through the corner points. 
Once the area has been delimited, the expert selects a category to which the new polygon should be added.
Then, the new polygon is automatically processed, intersected with the other polygons, and saved in the database.
Finally, through the linkage of the views, the information panel on the right lists the other polygons with which the newly created polygon intersects and which documents or restrictions apply to them. 
Moreover, if the polygon is selected, it is also displayed with an orange border on all other maps, making it clear where it is located in the shown area through brushing (\textbf{R3}).

\begin{figure}[tb]
	\centering
	    \begin{overpic}[trim={0 0 2.5cm 0},clip, width=\linewidth]{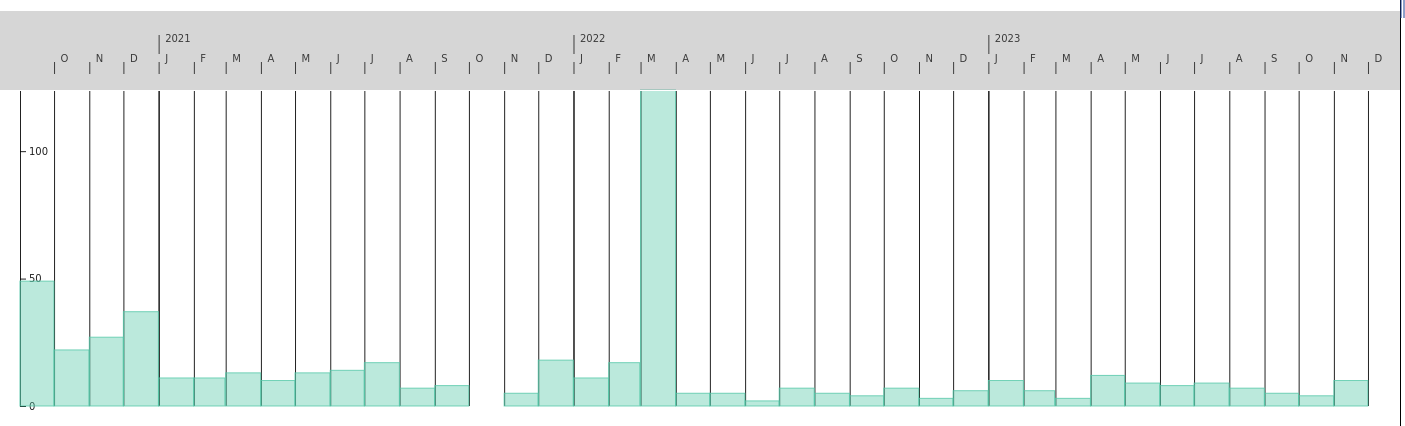}
     	 \put(-4,26){\textbf{\textit{a)}}}
     \end{overpic} 
         \begin{overpic}[width=\linewidth]{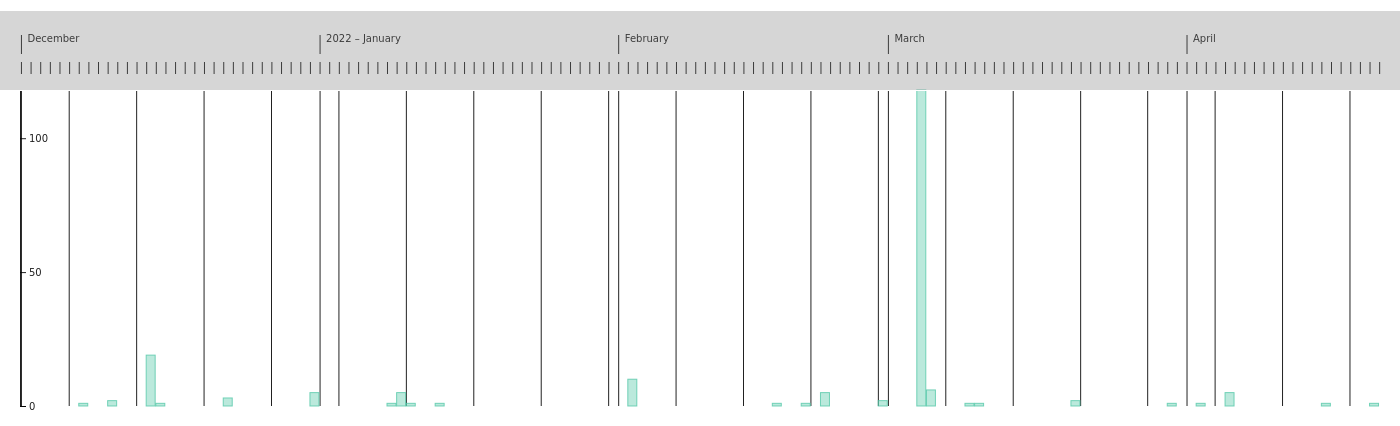}
     	 \put(-4,26){\textbf{\textit{b)}}}
     \end{overpic}  
     %\vspace{1cm}
    \begin{overpic}[width=\linewidth]{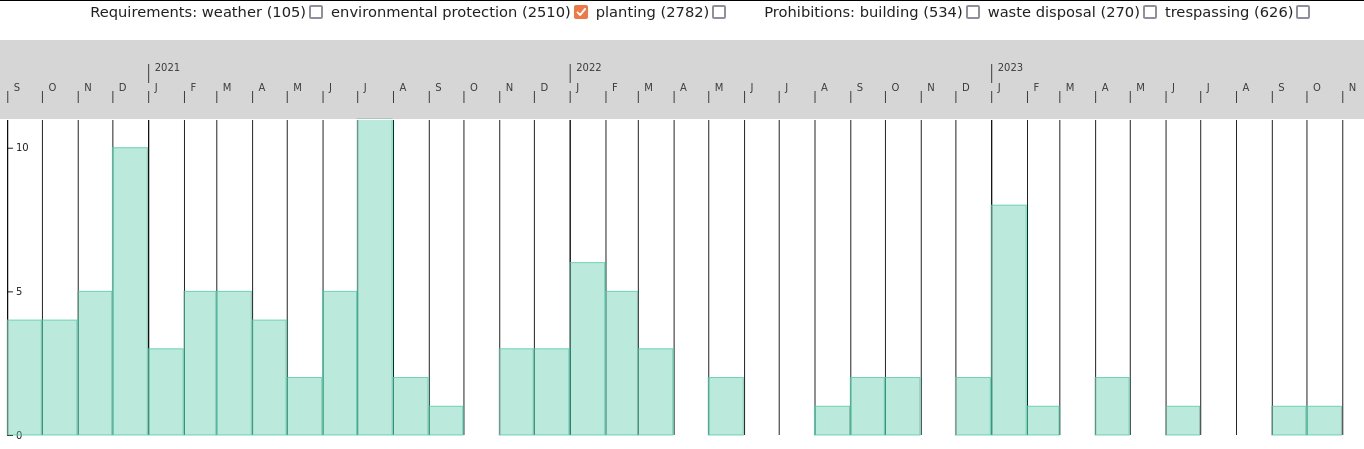} 
     	 \put(-4,26){\textbf{\textit{c)}}}
     \end{overpic}
	\caption[Component]{\label{fig:timeline}
		 The timeline visualization with different level-of-details and applied filters. (a) shows several months in between 2020 and 2023. (b) shows a part of the same interval on a daily-level, revealing that almost all of the documents for March 2022 are related to the 4th March. (c) shows the same interval as (a) when using the ``Requirement'' \textit{environmental protection} as mandatory.
	}
\end{figure}

%TODO  It would improve readability and clarity if, for example, lists of functions were presented in a table or list rather than in the body text (e.g. lines 289-304).
\subsection{Timeline}
% Design + interaction
The timeline view in Figure~\ref{fig:timeline} shows the number of documents associated with a specific date through green bars.
On the x-axis, the time range is displayed, and the width of a bar changes based on the provided space and the current level of detail.
On the y-axis, the number of documents is mapped on a linear scale between 0 and the highest number in the data set.
Based on the selected level of detail, the dates in the documents are counted for each bar.
The visualization allows for the selection of a specific bar and brushing of a time interval.
This allows to explore all documents associated to the corresponding time range in an information panel and to view the associated polygons on a map.

\begin{figure}[h!]
	\centering
	    \begin{overpic}[trim={0 0 0 0},clip, width=\linewidth]{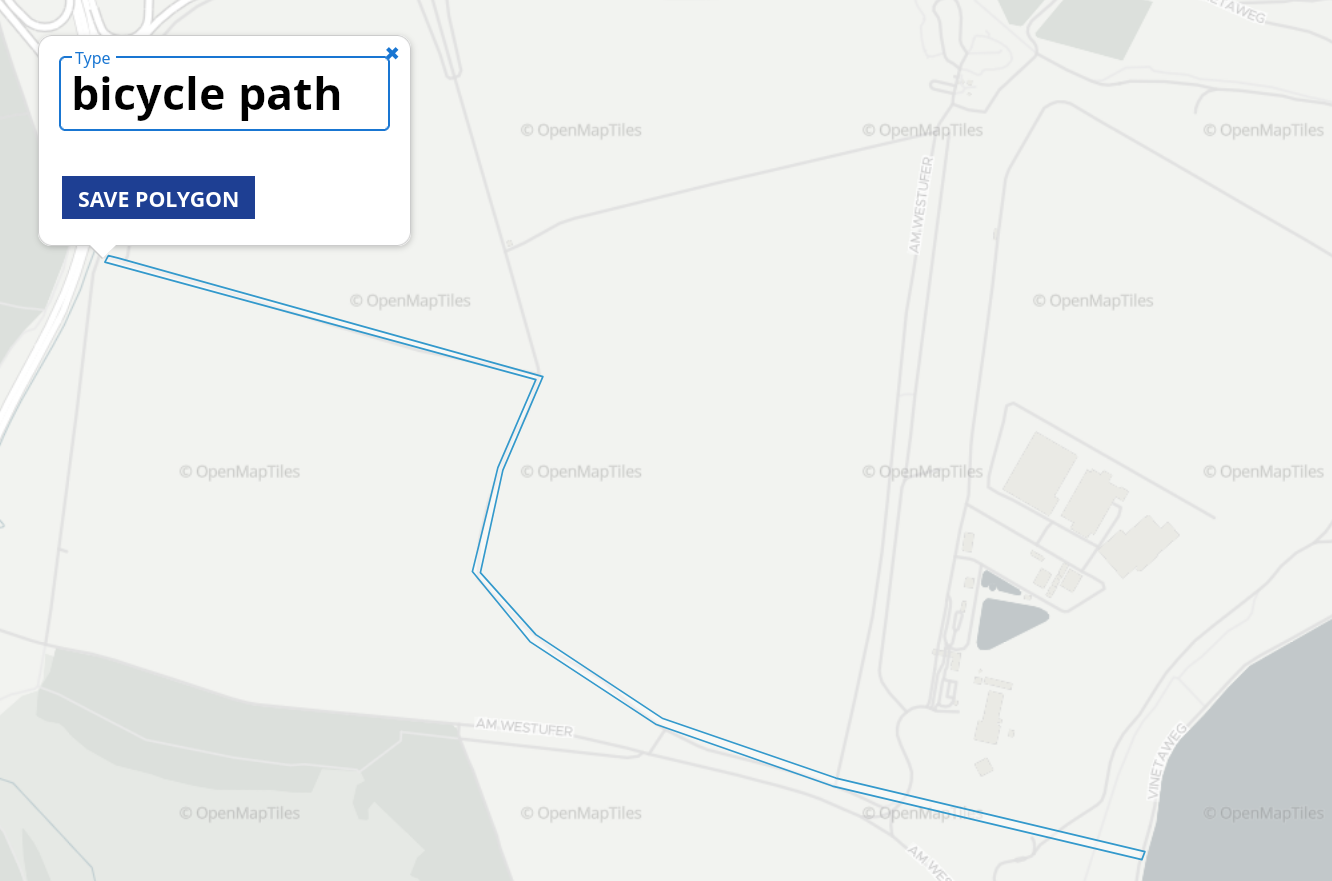}
     	    \put(-4,64){\textbf{\textit{a)}}}
        \end{overpic} 
         \begin{overpic}[trim={2 2 2 0},clip, width=\linewidth]{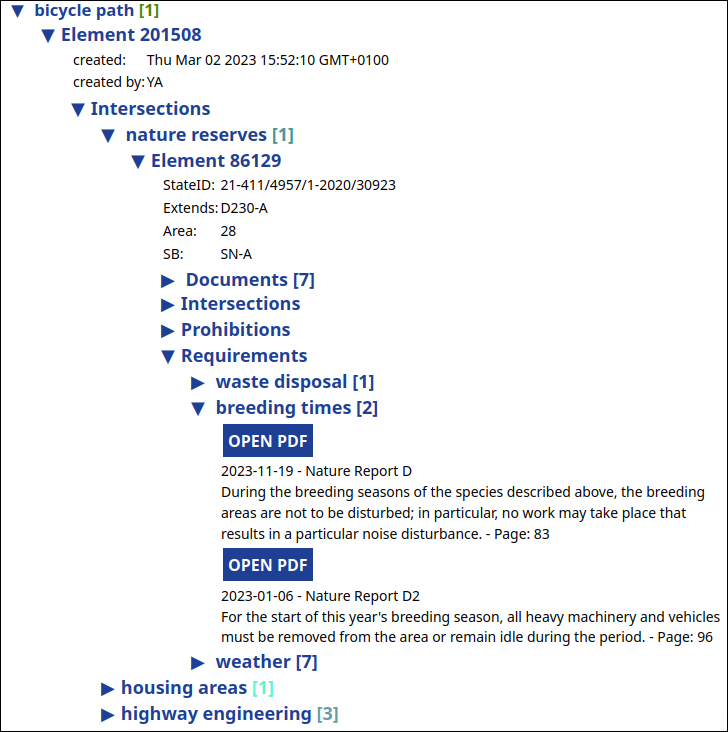}
     	     \put(-4,97){\textbf{\textit{b)}}}
        \end{overpic}  
	\caption[Component]{\label{fig:usecasePath}
		 (a) A new project, a bicycle path, is drawn over the map and then saved in the database.
		 (b) Then other projects are listed in the information panel that overlap with the newly created one. 
		 Here, a nature reserve is listed, which is subject to various restrictions, including two from the category of breeding times.
	}
\end{figure}

It is also possible to filter the timeline on the basis of the current viewport of a map view through the linkage of the views.
This allows a domain expert to focus on one area of interest when planning new projects.
Furthermore, the timeline supports zooming into a specific time range, thus changing the current level of detail.
Based on the zoom level, the bars in the timeline represent a decade, year, quarter, months, or a day.
Figure~\ref{fig:timeline}a-b shows examples for months and days.
This enables a domain expert to focus on high-level patterns, such as yearly or monthly restrictions, or to look into a specific day of interest.
It is also possible to filter the timeline based on a time range, restrictions, or to combine these filters with the viewport filter.
The filtering of specific restrictions eases the decision-making processes by showing only the relevant time data for a specific domain question, and thus supports \textbf{R1}.
Figure~\ref{fig:timeline}c shows a filtered version of the timeline in Figure~\ref{fig:timeline}a for the restriction \textit{environmental protection}.

\section{Use Case}
\begin{figure*}[tb]
	\centering
	    \begin{overpic}[trim={0 0 0 0},clip, width=\linewidth]{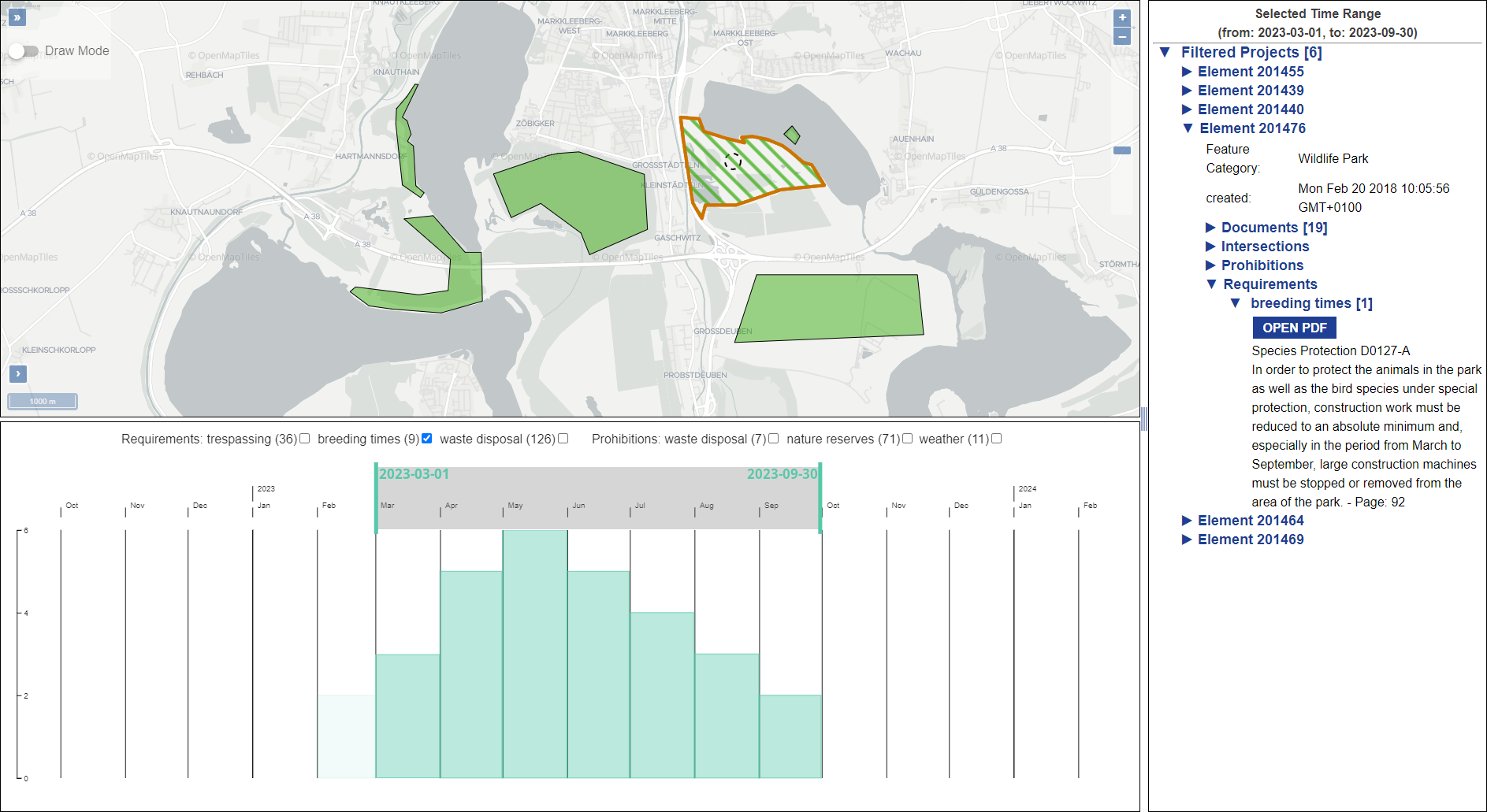}
     	    \put(2,52.9){\textbf{\textit{a)}}}
     	    \put(78,52.9){\textbf{\textit{b)}}}
     	    \put(2,24.7){\textbf{\textit{c)}}}
        \end{overpic} 
	\caption[Component]{\label{fig:usecase2}
		 (c) In the timeline view, the documents are filtered according to the displayed region in the map and the Requirement ``breeding times''. The time range from March to September was selected.
		 (a) Through brushing, the associated projects of the selected range are displayed on the map.
		 (b) They are additionally listed with further information in the information panel.
		 A selected project states that certain noisy activities are apparently not permitted in the associated wildlife park during the specified time.
	}
\end{figure*}

Refer to the previously described task, where a new project is to be planned along existing ones. 
%In this use case, a bicycle path is to be planned in the region around Markkleeberg (south of Leipzig). 
This use case is about the planning of a bicycle path in the region around Markkleeberg, which is south of Leipzig. 
The area is also known as the ``Leipziger Neuseenland'' (New Lakeland) and was created by the renaturation of several open-cast mining areas.
As a result, there are already other existing projects along the planned route that are subject to certain restrictions. 
%Along the planned route, there are already other projects that are subject to certain restrictions. 
If their area overlaps with that of the planned path, these would also apply to the new project. 
To find out which existing restrictions could be relevant and which are already known, an expert simply draws the bicycle path to be planned on the map in our GIS tool (\textbf{R2}, \textbf{R3}, Figure \ref{fig:usecasePath}a).

Once the path has been planned, the expert selects a category in which the path should be saved, and then our tool calculates all overlaps with other areas in this area. 
Due to the structure in the database, it only takes a few seconds until the result is displayed in the information panel on the right. 
All results listed there are areas from other projects that intersect the bicycle path.
From these areas, the already known restrictions to the associated projects are listed (\textbf{R3}).

For example, there are restrictions on nature conservation, for instance, some areas may not be entered or damaged, and building waste may not be stored there. 
However, in particular, there are also restrictions on the subject of breeding times (\textbf{R1}, Figure \ref{fig:usecasePath}b).
The expert now knows that in some months the construction project may only take place under special conditions or, in some cases, may not take place at all, as noise would disturb the animals, and construction vehicles are thus also not allowed to be active there.
With this knowledge, the expert can now better time the path and the aspects necessary for it.

Once planned, the project is not finished yet. 
Further planning is necessary, and projects need to be monitored for compliance with certain regulations, including temporal restrictions (\textbf{R1}).
For this purpose, existing projects can be displayed (see Figure \ref{fig:usecase2}).
To do this, an expert centers the map in a specific region.
Then, the expert displays the timeline and filters the projects listed there according to the centered region and subsequently to the breeding times category (Figure \ref{fig:usecase2}c). 
In this way, the tool filters all existing projects accordingly and shows which projects and especially when these projects are subject to time restrictions on breeding times.
The example shows a project that has to observe breeding times from March to September (Figure \ref{fig:usecase2}b).
As a consequence, large construction vehicles may not be used in the area, for example.

This and also other projects that match the filter are displayed on the map to get a better overview of exactly which areas are involved (Figure \ref{fig:usecase2}a).
With this information, the expert can now plan the exact resources of these projects, so that construction machines and vehicles can be used elsewhere, as they are not allowed here at this time anyway. 
The planning is thus more precise and efficient.
However, the expert can also check whether these projects adhere to the restrictions and initiate appropriate corrective actions.
\section{Limitations \& Future Work}
The presented system can support decision-making in urban planning, renaturation, and redevelopment projects, but there are still some limitations that can be addressed in the future. 
Sentences regarding breeding times can be found well, but some of them lack time specifications.
Often these are not given directly in the sentence but can be obtained in the context of the paragraph, or they can also result from the named bird species, as these always breed at certain times, and thus the time specifications are not explicitly given. 
In all these cases, the extracted data had to be manually linked to their temporal aspects. 
A major simplification of this work can be achieved by adapting the information extraction to this and examining the context for exactly such temporal aspects.
Furthermore, the including of uncertainties based on the extraction process and the correction, or labeling of entries could be helpful for the domain experts.

So far, we have only referred to bird breeding times.
However, the tool could also be extended to other categories, such as plant pruning times, flowering times, and pollen flight times.
Currently, we only focus on a subset of documents in the Lausitzer und Mitteldeutsche region, by combining different data sources and using data from different regions, the system could help other communities in regional planning.

\section{Conclusion}
We present a system to support decision-making in urban planning, renaturation, and redevelopment projects.
Legal restrictions and other limitations were extracted from georeferenced documents and linked to their temporal facet.
This means that it is no longer necessary to manually sift through vast quantities of text documents in search of these restrictions.
The tool allows to create new polygons to plan new projects, such as bicycle path, while considering spatial and temporal restrictions (``Prohibitions'' and ``Requirements''), like bird breeding or tree pruning periods.  

For this, a map is linked to a timeline visualization.
By using this timeline, existing projects can be filtered and found according to the temporal component. 
This makes it easier to plan when and where resources are needed for specific projects.
With this, our presented system can also be used to verify compliance with certain legal and nature conservation regulations.
In summary, the use cases show that the tasks posed at the beginning can be accomplished with our system.
In addition, existing knowledge no longer has to rely solely on the memory of certain experts, but this knowledge has been made explicit and discoverable for all, 
  making the system useful for our project partner, the LMBV.

\section*{Acknowledgment}
This research was supported by the Development Bank of Saxony (SAB) under Grant 100400221.
We thank our project partner the Lausitzer und Mitteldeutsche Bergbauverwaltungsgesellschaft mbh (LMBV) for their support and for providing their data.

% Acknowledgment?
%-------------------------------------------------------------------------

%\bibliographystyle{eg-alpha}
\bibliographystyle{eg-alpha-doi}

\bibliography{envirvis2023}

%-------------------------------------------------------------------------
\newpage

\end{document}